# Accelerator and Beam Physics Research Goals and Opportunities

**Working group:** S. Nagaitsev (Fermilab/U.Chicago) *Chair*, Z. Huang (SLAC/Stanford), J. Power (ANL), J.-L. Vay (LBNL), P. Piot (NIU/ANL), L. Spentzouris (IIT), and J. Rosenzweig (UCLA)

**Workshops conveners:** Y. Cai (SLAC), S. Cousineau (ORNL/UT), M. Conde (ANL), M. Hogan (SLAC), A. Valishev (Fermilab), M. Minty (BNL), T. Zolkin (Fermilab), X. Huang (ANL), V. Shiltsev (Fermilab), J. Seeman (SLAC), J. Byrd (ANL), and Y. Hao (MSU/FRIB)

**Advisors:** B. Dunham (SLAC), B. Carlsten (LANL), A. Seryi (JLab), and R. Patterson (Cornell)

*January 2021*

# Abbreviations and Acronyms

| | |
|---|---|
| **2D** | two-dimensional |
| **3D** | three-dimensional |
| **4D** | four-dimensional |
| **6D** | six-dimensional |
| **AAC** | Advanced Accelerator Concepts |
| **ABP** | Accelerator and Beam Physics |
| **DOE** | Department of Energy |
| **FEL** | Free-Electron Laser |
| **GARD** | General Accelerator R&D |
| **GC** | Grand Challenge |
| **H-** | a negatively charged Hydrogen ion |
| **HEP** | High-Energy Physics |
| **HEPAP** | High-Energy Physics Advisory Panel |
| **HFM** | High-Field Magnets |
| **KV** | Kapchinsky-Vladimirsky (distribution) |
| **ML/AI** | Machine Learning/Artificial Intelligence |
| **NCRF** | Normal-Conducting Radio-Frequency |
| **NNSA** | National Nuclear Security Administration |
| **NSF** | National Science Foundation |
| **OHEP** | Office of High Energy Physics |
| **QED** | Quantum Electrodynamics |
| **rf** | radio-frequency |
| **RMS** | Root Mean Square |
| **SCRF** | Super-Conducting Radio-Frequency |
| **USPAS** | US Particle Accelerator School |
| **WG** | Working Group |

# Accelerator and Beam Physics

1. **EXECUTIVE SUMMARY**

Accelerators are a key capability for enabling discoveries in many fields such as Elementary Particle Physics, Nuclear Physics, and Materials Sciences. While recognizing the past dramatic successes of accelerator-based particle physics research, the April 2015 report of the Accelerator Research and Development Subpanel of HEPAP [1] recommended the development of a long-term vision and a roadmap for accelerator science and technology to enable future DOE HEP capabilities.

The present report is a summary of two preparatory workshops, documenting the community vision for the national accelerator and beam physics research program. It identifies the Grand Challenges of accelerator and beam physics (ABP) field and documents research opportunities to address these Grand Challenges. This report will be used to develop a strategic research roadmap for the field of accelerator science.

Four accelerator and beam physics grand challenges were identified:

**Grand Challenge #1:** Beam Intensity – "How do we increase beam intensities by orders of magnitude?"

**Grand Challenge #2:** Beam Quality – "How do we increase the beam phase space density by an order of magnitude, towards the quantum degeneracy limit?"

**Grand Challenge #3:** Beam Control – "How do we measure and control the beam distribution down to the individual particle level?"

**Grand Challenge #4:** Beam Prediction – "How do we develop predictive 'virtual particle accelerators'?"

The ABP research community input during the two workshops indicated the following areas of research are needed to address the above Grand Challenges:

| **Proposed research area** | GC1 | GC2 | GC3 | GC4 |
|---|---|---|---|---|
| Single-particle dynamics and non-linear phenomena. Polarized beams dynamics | A | A | B | B |
| Collective effects (e.g. space charge, beam-beam, and self-interaction via radiative fields, e.g., coherent synchrotron radiation) and mitigation | A | A | B | B |
| Beam instabilities, control, and mitigation; (e.g. short- and long-range wakefields) | A | A | B | B |
| High-brightness beam generation: low emittance high-peak current, and ultrashort bunches | | A | B | B |
| Beam quality preservation and advanced beam manipulations. Beam cooling and radiation effects in beam dynamics | | A | B | B |
| Advanced accelerator instrumentation and controls | | | A | |
| High-performance computing algorithms, modeling and simulation tools | | | | A |
| Fundamental accelerator theory and applied math | A | A | A | A |
| Machine learning and artificial intelligence | A | A | A | A |
| Early integration, optimization, and maturity evaluation of accelerator and project concepts (focuses on science and technology gaps and bridges between various R&D thrusts and particle physics requirements) | B | B | B | B |

A = research **addresses** (targets) this GC, B = research **benefits** from addressing this GC

It was noted during the WG7 workshop sessions (see Section 5.3) that the HEP ABP research program is the most appropriate and capable of providing systematic support to the general US accelerator community in its "Early Conceptual Integration and Optimization, Maturity Evaluation" efforts.

## 2. INTRODUCTION

The Accelerator and Beam Physics (ABP) thrust is part of the DOE HEP-funded research portfolio, focused on General Accelerator R&D (GARD). A working group, consisting of S. Nagaitsev (Fermilab/UChicago), Z. Huang (SLAC/Stanford) *Chair*, J. Power (ANL), J.-L. Vay (LBNL), P. Piot (NIU/ANL), L. Spentzouris (IIT), and J. Rosenzweig (UCLA) was established in September 2018 to develop a community-driven research roadmap. A series of working group meetings and community workshops [2, 3] were conducted in 2019-2020 to gather input for this document.

**Definition**

Accelerator and beam physics is the science of the motion, generation, acceleration, manipulation, prediction, observation and use of charged particle beams. The Accelerator and Beam Physics (ABP) thrust focuses on fundamental long-term accelerator and beam physics research and development.

**Vision statement**

Particle accelerators can be used to better understand our universe and to aid in solving societal challenges. The ABP thrust explores and develops the science of accelerators and beams to make future accelerators better, cheaper, safer, and more reliable.

**ABP missions**

The primary scientific mission of the ABP thrust is to address and resolve the Accelerator and Beam Physics Grand Challenges, outlined below. Other equally important ABP missions are associated with the overall DOE HEP missions:

1. Advance the physics of accelerators and beams to enable future accelerators.
2. Develop conventional and advanced accelerator concepts and tools to disrupt existing costly technology paradigms in coordination with other GARD thrusts.
3. Guide and help to fully exploit science at the HEP GARD beam facilities and operational accelerators.
4. Educate and train future accelerator scientists and engineers.

**3. ACCELERATOR and BEAM PHYSICS GRAND CHALLENGES**

**3.1. Grand Challenge #1: Beam Intensity -- "How do we increase beam intensities by orders of magnitude?"**

Beam intensities in existing accelerators are limited by collective effects and particle losses. A complete and robust understanding of these effects is necessary to help overcome the limits and increase beam intensities by orders of magnitude. Grand Challenge #1 addresses the question: "How do we increase beam intensities by orders of magnitude?"

**Description**

Future demands for beams will exceed present capabilities by at least an order of magnitude in several parameter regimes such as average beam power and peak beam intensity. Ultimately, the beam intensities attainable in present accelerators are limited by collective effects and particle losses from various sources, e.g. space-charge forces, beam instabilities, and beam injection losses. How do we overcome the collective forces in the beam that deteriorate beam properties and lead to beam losses? A robust understanding of these effects in real accelerators does not yet exist. Additionally, theoretical, computational, and instrumentation tools to address this challenge are not yet fully developed at the precision level required by modern beam applications (see GC #3 and #4).

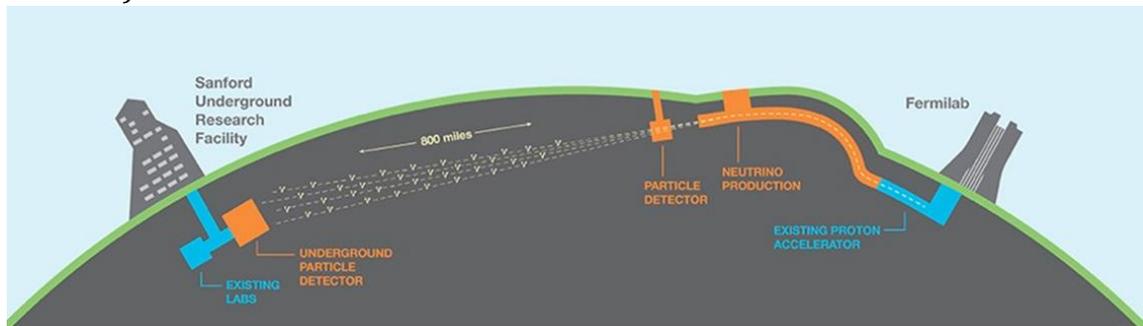

Addressing GC1 would enable most intense neutrino beams (image courtesy of Fermilab).

**Scientific impacts and dividends**

- Deliver an order of magnitude increase or more in secondary particle fluxes from proton and heavy-ion driver applications;
- Improve performance of beam-driven wakefield accelerators;
- Enable ultrashort electron bunches for collider applications;
- Enable first generation of accelerator-driven energy systems;
- Inform challenges associated with beam quality, control and prediction.

**3.2. Grand Challenge #2: Beam Quality -- "How do we increase the beam phase space density by order of magnitude, towards the quantum-degeneracy limit?"**

Most applications of accelerators depend critically on the beam intensity and directionality (or beam emittance), in order to enable new capabilities or to optimize the signal to noise ratio. Addressing this grand challenge will yield unprecedented beam qualities that can revolutionize applications of particle accelerators. Grand Challenge #2 addresses the question: "How do we increase the beam phase space density by order of magnitude, towards the quantum degeneracy limit?"

**Description**

The beam phase-space density is a determining factor for the luminosity of high-energy colliders, for the brightness of photon sources based on storage rings or free-electron lasers (FELs), and on emergent instruments using electrons to probe matter — ultra-relativistic electron diffraction and microscopy. Pushing beam phase-space density beyond the current state-of-the-art has tremendous payoffs for discovery sciences driven by accelerators. It will permit FELs with new capabilities, enable femtosecond to attosecond resolution electron imaging, and provide new tools for future collider development through source and wakefield accelerator research. Research topics span frontier schemes for generating high-brightness electron and proton/hadron beams; controlling space charge and coherent radiation effects and other collective instabilities; preserving beam brightness during beam generation and acceleration, compression and manipulations; and developing novel techniques for beam cooling to increase phase-space density.

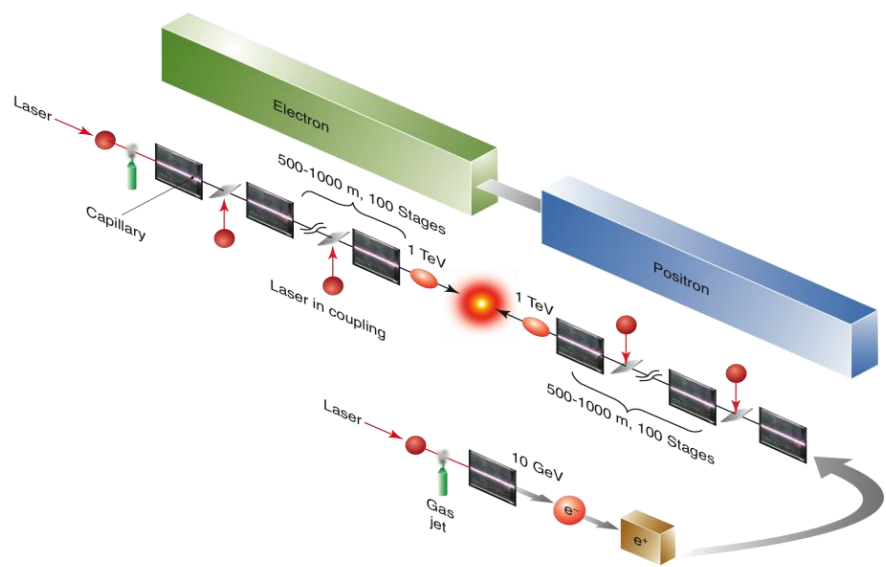

Addressing GC2 would enable high-luminosity colliders (image reproduced from Leemans & Esarey, Physics Today, 2009).

**Scientific impacts and dividends**

- Create new paths for dramatically increased collider luminosity;
- Enable compact wakefield-based colliders;
- Significantly enhance the brightness and wavelength reach of modern X-ray sources;
- Enable schemes for compact FELs;

- Create beam-based tools with unprecedented temporal and spatial resolution.

## 3.3. Grand Challenge #3: Beam Control -- "How do we measure and control the beam distribution down to the individual particle level?"

An accelerator application benefits most when the beam distribution is specifically matched to that application. This challenge aims to replace traditional methods that use beams of limited shapes with new methods that generate tailored beams. It also aims to provide new research opportunities, enabled by detecting and controlling individual particles in accelerators and storage rings. Grand Challenge #3 addresses the question: "How do we measure and control the beam distribution down to the individual particle level?"

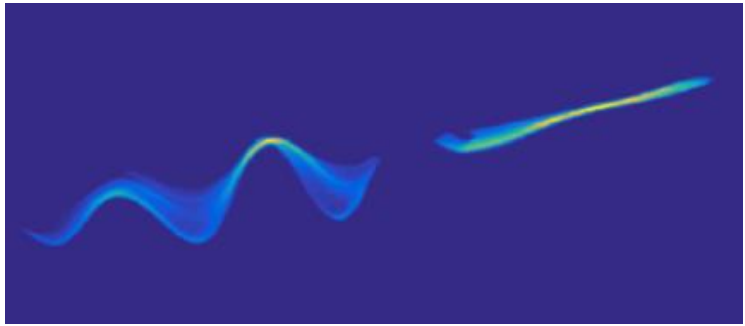

A single-shot longitudinal phase-space measurement in the presence of beam wakefields (image courtesy of ANL).

**Description**

A given accelerator application is best served by a beam with a specific distribution, matched to the application. These applications may include, for example, extra-short electron bunches from photo-cathodes for colliders or proton injection painting to mitigate losses in high-intensity synchrotrons. The goal of controlling and creating specific beam distributions at a fine level represents a paradigm shift from traditional approaches, based on rms or higher-level beam properties. This new approach presents significant challenges in beam dynamics and diagnostics as well as in accelerator design and operation. This grand challenge seeks to develop techniques, beam diagnostics, and beam collimation methods with the ultimate goal of controlling the complete 6D phase space distribution at the individual particle level and near-term milestones capable of controlling and detecting the beam distribution towards the attosecond time scale and nanometer spatial scale. An associated objective is to develop techniques to track and control individual particles or pre-formatted groups of particles. In addition, given the complexity of these 6D distributions and the associated collective effects, the use of machine learning and artificial intelligence (ML/AI) to control the beam distribution (in simulations or during accelerator operation) should be explored.

**Scientific impacts and dividends**

- Substantially increase luminosity in future colliders;
- Mitigate beam losses;
- Improve the performance of future advanced collider concepts;
- Enable table-top coherent light sources;
- Enable quantum science experiments.

### 3.4. Grand Challenge #4: Beam Prediction -- "How do we develop predictive 'virtual particle accelerators'?"

Developing "virtual particle accelerators" will provide predictive tools that enable fast computer modeling of particle beams and accelerators at unprecedented levels of accuracy and completeness. These tools will enable or speed up the realization of beams of extreme intensity and quality, as well as enabling control of the beam distribution reaching down to the level of individual particles. Grand Challenge #4 addresses the question: "How do we develop predictive 'virtual particle accelerators'?"

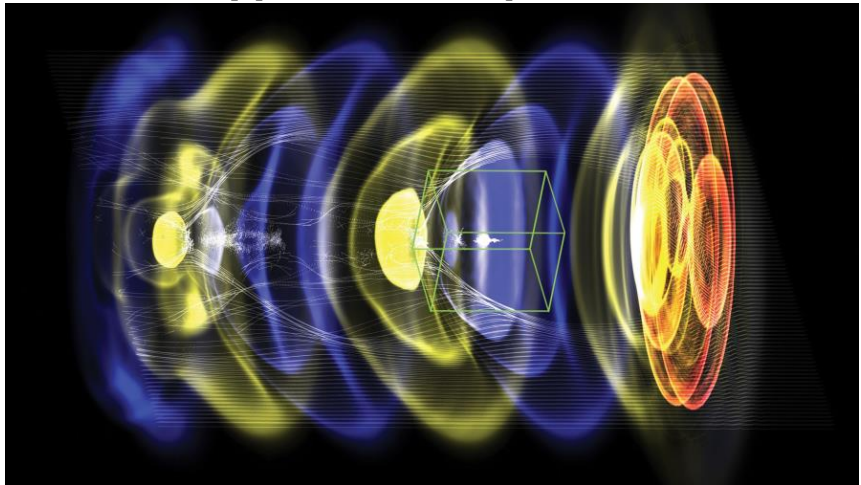

A laser-driven plasma accelerator stage: a rendering from a 3D WarpX simulation with mesh refinement (image courtesy of LBNL).

**Description**

The importance of particle accelerators to society, along with their increasing complexity and the high cost of new accelerator facilities, demand that the most advanced computing and ML/AI tools be brought to bear on R&D activities in particle beam and accelerator science. Pushing the limits in beam intensity, quality and control demands more accurate, more complete and faster predictive tools, with an ultimate goal of virtual accelerators. The development of such tools requires continuous advances in fundamental beam theory and applied mathematics, improvements in mathematical formulations and algorithms, and their optimized implementation on the latest computer architectures. The modeling of beams at extreme intensities and levels of quality, and the design of accelerators that deliver them, also call for integrated predictive tools that can take advantage of high-performance computing. Full integration of machine learning tools (and their further development when needed) will be essential to speed up the realization and boost the power of virtual particle accelerators.

**Scientific impacts and dividends**

- Deliver an integrated ecosystem of predictive tools for accurate, complete and fast modeling of particle accelerators and beams;
- Enable virtual accelerators that can predict the behavior of particle beams in accelerators "as designed" or "as built";
- Provide the predictive tools that will enable or speed up the realization of the beam intensity, quality, and control grand challenges;
- Develop mathematical and algorithmic tools that benefit from — and contribute to — synergistic developments beyond particle beam and accelerator science.;
- Maximize the benefits from - and to – ML/AI tools for beam science and accelerator design.

4. **SUMMARY from WORKSHOP #1**

**HEP GARD Accelerator and Beam Physics: Community-driven Strategic Roadmap Workshop #1 (LBNL, Dec 9-10, 2019)**

**Workshop #1 goals**

The goal of the workshop [2] was to identify the key Accelerator and Beam Physics (ABP) needs to fulfill OHEP's GARD mission and to develop a decadal roadmap for thrust activities for OHEP to support.

Workshop #1 focused on research areas for each of four working groups:

**WG1:** Single-particle dynamics, including nonlinearities, and spin dynamics.
**WG2:** High-brightness beam generation (including polarized beams), transport, manipulation and cooling.
**WG3:** Mitigation and control of collective phenomena: instabilities, space charge, beam-beam, beam-ion effects, wakefields, and coherent synchrotron radiation.
**WG4:** Connections to other GARD roadmaps (cross-cutting WG1-3)

## 4.1. WG1: Single particle dynamics, including non-linearities and spin dynamics

The themes in this section are: (1) advancement in new analytic methods; (2) strong transverse focusing in colliders and storage rings; and (3), beam focusing schemes with reduced chaos and beam loss.

The recommendations from all of the themes in WG1 are:
- Strive for greater theoretical understanding of underpinning dynamical phenomena;
- Use operational and R&D test facilities as a bridge from theory to experimental tests. Use small size dedicated R&D machines for proof-of-principle research. Use production machines for large scale tests.
- Realize virtual accelerators (connection to Workshop 2)
- Investigate how the efficiency and cost of accelerators are related to beam physics

### 4.1.1. Theme 1: Advancement in new analytic methods

New analytic and theoretical methods to construct a 4D or 6D time-independent Hamiltonian are required in order to advance our understanding of single particle dynamics in future accelerators. These theoretical advances would help address all four Grand Challenges as they would form the underpinning of next generation beam dynamics. New beam dynamics methods would include the analysis of parasitic and intentional nonlinearities in a ring, dynamic aperture predictions and optimization, new high-efficiency slow extraction methods and investigation on non-perturbative nonlinear optics schemes.

**Research topics in Theme 1:**

Long-term dynamic aperture in storage rings
Currently, the reliable tracking of dynamic aperture in storage rings, such as the Large Hadron Collider, spans about 10 million turns. However, it is highly desirable to track up to $10^8$-$10^9$ turns. What is the ultimate limit of long-term tracking in storage rings? What are the limiting factors? How do tracking results compare to measurements, including at ultimate tracking limits?

Better understanding of dynamic aperture
Currently, our understanding of dynamic aperture is mostly based on nonlinear resonances, tune shifts, and chromaticity as derived from conventional perturbation theory. A detailed knowledge of how these affect the dynamic aperture in storage rings is needed. Can the effective Hamiltonian be used to improve understanding? Is there any relationship between periodic orbits and dynamic aperture?

Optimization of dynamic aperture
It is essential to develop a systematic and robust scheme to optimize the dynamic aperture, not only for synchrotron light sources with a large super periodicity, but also for the high-energy colliders. The optimization should rely on a solid understanding of dynamic aperture and benchmarked against direct tracking results or measurements. Machine learning algorithms should be explored in the development of computational tools.

High-efficiency slow extraction methods
Improvements to slow extraction methods and systems for high efficiency and high beam power would be an important enhancement to machine capability for particle physics at the intensity frontier.

### 4.1.2. Theme 2: Strong transverse focusing in colliders and storage rings

Prior research into strong transverse focusing (small average beta-functions) in storage rings and colliders indicates that it could mitigate intrabeam scattering and other undesirable effects. However, strong focusing also produces strong chromatic effects. Combining strong focusing with chromatic corrections is a very promising area of research to address Grand Challenges #1 and #2. This advancement would require increased precision in measurement capability and in control of optics (~1% for beta-functions) to achieve the needed tunability of accelerators.

**Research topics in Theme 2:**

Mitigation of rapidly increasing nonlinearity in storage rings
To further increase the luminosity of colliders or the brightness of synchrotron light sources, the beam emittance and beam size are being continually pushed downward, recently to the picometer region in electron rings. In the next decade, another order of magnitude reduction is expected. As a result of ever smaller emittance, chromatic sextupoles become increasingly stronger. Can the mitigation of nonlinearities be systematically developed to reduce the nonlinear effects of the storage rings?

Correction of chromatic effects in advanced linear collider schemes
All advanced linear collider schemes should be adaptable to beams with higher than normal momentum spreads. High momentum spreads may arise from beam loading but may also be needed for damping of transverse oscillations and to mitigate the beam-break-up instability. Transporting and focusing of beams with high momentum spread would open up many opportunities in advanced accelerator schemes.

### 4.1.3. Theme 3: Beam focusing schemes with reduced chaos and beam loss

Nonlinear optics (transverse focusing) schemes in high-intensity circular accelerators would enhance a beam's immunity to instabilities and losses.

**Research topics in Theme 3:**
<u>Nonlinear transverse focusing</u>
Nonlinear transverse focusing schemes with reduced chaos hold great promise for enhanced Landau damping and instability suppression by tune amplitude-dependent tune spread.

<u>IOTA facility research plan</u>
The experimental research plan at the IOTA facility progresses from control of single particle amplitude-dependent tune-shift to Landau damping of intense beams.

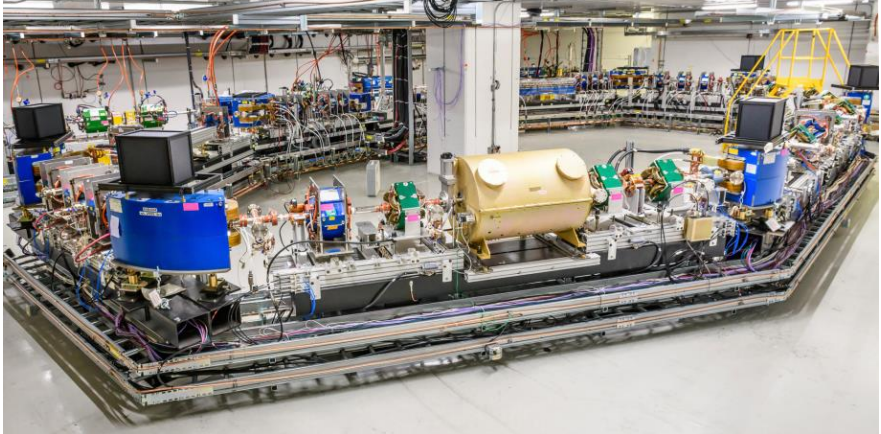

Integrable Optics Test Accelerator (IOTA), a versatile particle storage ring facility at Fermilab (photo: Giulio Stancari, Fermilab).

## 4.2. WG2: High-brightness beam generation, transport, manipulation and cooling

WG2 focused on research needs and opportunities in beam generation, transport, manipulation, and cooling. Such topics are relevant to the four grand challenges and are connected to the energy and intensity frontiers along with other accelerator applications beyond high-energy physics. The main themes in this section are: (1) controlling the beam phase-space distribution; and (2), decreasing the phase-space volume.

### 4.2.1. Theme 1: Controlling the beam phase-space distribution

Research into how the beam distributions (or RMS parameters) can be tailored to match a specific front-end application or mitigate collective effects.

**Research topics and milestones in Theme 1:**

<u>Further development of beam manipulation models and methods</u>
Developing beam manipulation initially with one degree of freedom (e.g. within the longitudinal phase space), but also aiming at 3D beam manipulation in the future. The specific example discussed was the generation of MegaAmp-class peak-current electron bunches with potential applications as QED and extreme physics probes. The approach relies on conventional magnetic bunch compression with associated limitations coming from collective effects (radiative effects such as coherent synchrotron radiation, and space charge). Understanding, modeling, and devising mitigation techniques to control these collective effects is an important aspect of bunch compression.

Designing new techniques for phase space manipulation
Designing new techniques to shape the phase space of a beam along any degree of freedom. One research direction concerns the development of phase-space manipulation based on an exchange between two degrees of freedom (demonstrated for e-beam but in principle applicable to other particle species). Another research topic concerns the development of advanced phase-space painting techniques to produce a beam with linearized space charge forces (i.e. KV-like distribution) in rings.

Controlling a beam using the field produced by another co-propagating or counter propagating beam; for example, one of the methods discussed was the use of an electron lens to collimate the halo associated with a proton beam. A variant of the electron-lens concept could in principle be expanded to enable more precise control over the distribution of a hadron beam. Some of the related research topics include mitigation of space-charge effects, control of tune shift, and "enhanced" Landau damping.

Using external electromagnetic fields to control the beam; for example, the technique of laser-based stripping as a method to produce high intensity structured proton beams using charge exchange from H-. The method overcomes limitations associated with foil-based charge exchange techniques and could enable high-intensity proton beams. It presents some research challenges in the development of high-power pulsed lasers.

### 4.2.2. Theme 2: Decreasing the phase-space volume

From beam generation to the final application, the beam brightness can at best be conserved without employing cooling methods. Cooling could provide opportunities to reduce the beam phase space(s). Several advanced cooling techniques were discussed.

**Research topics and milestones in Theme 2:**

Micro-bunched electron cooling methods
Micro-bunched electron cooling methods include techniques such as encoding the hadron beam information on an electron beam and then amplifying it via space charge before interacting back with the hadron beam. This cooling technique presents challenges in the generation of cold low-noise electron beams.

Optical stochastic cooling techniques
Optical stochastic cooling techniques are applicable to any beam species and present a number of research opportunities related to the precise control of the optical field and beam distribution.

Laser-based cooling
Laser-based cooling is capable of attaining mK to µK-level temperatures in the beam frame, and could, in principle, cool the beam in all phase spaces and to achieve quantum limits.

### 4.3. WG3: Mitigation and control of collective phenomena

WG3 focused on the understanding and control of collective effects, which is a theme most closely related to Intensity Frontier challenges, with occasional relevance to Energy Frontier challenges. Each theme in this WG is related to at least one grand challenge. The themes in this section are: (1) space charge compensation schemes; (2) quantitative predictive power from simulations; (3) advancement in beam diagnostic techniques; and (4), test stands to advance new technology and validate simulations.

The recommendations from all of the themes in WG1 are:
Strive for greater theoretical understanding of underpinning dynamical phenomena;

#### 4.3.1. Theme 1: Space charge compensation schemes

**Research topics in Theme 1:**

Electron lenses
Electron lenses provide a type of compensation with a broad range of applications. Electron lenses have previously been tested in some machines but require evolution of their capabilities to address the needs of next generation facilities. A particularly promising application is to use electron lenses for compensation of space charge effects and reduced tune depression. Such a scheme is currently under development for testing at the IOTA facility.

Other promising electron lens compensation schemes include electron lens collimation, Landau damping of proton instabilities, and compensation for beam-beam effects. Some lens schemes may be effective from a physics standpoint but impractical from a technological or cost standpoint due to required lengths or densities.

Relation of space charge effects to instabilities
Increasing intensity also requires control of instabilities. The impact of space charge on instabilities is not completely understood and requires progress both on predictive techniques (simulations and experiments), and mitigative methods.

### 4.3.2. Theme 2: Quantitative predictive power from simulations

**Research topics in Theme 2:**

Validating codes with data
For both linear accelerators and rings, simulations have evolved significantly over the past two decades and have been successful at modeling qualitative behavior in a beam. However, they have not yet reached the level of being able to make quantitative predictions in all cases. A more rigorous effort to experimentally benchmark simulations against measured data and understand code deficiencies is necessary to advance the predictive power of simulations required to reach the next decade in energy or intensity. Confidence in predictions for future machines relies on validating the codes with data from existing facilities and test accelerators.

High fidelity models
In the high intensity regime where beam loss is a central issue, small fractional quantities of beam near the edges of the distribution must be accurately modeled in order to gain a meaningful result from the simulation. Current status of benchmarking results is far from this goal. Advancements in modeling are also required to identify instability thresholds for both wakefields and e-cloud based instabilities in future accelerators.

The need for accurate simulations is common across all four grand challenges.

### 4.3.3. Theme 3: Advancement in beam diagnostic techniques

**Research topics in Theme 3:**

Better diagnostic interfacing between diagnostics and computational tools
More advanced diagnostics are needed in order to bridge the gap between experiment and simulation in several areas.

Performance improvement and development of innovative diagnostics
Predicting e-cloud relies on accurate benchmarking against experiment, however, instrumentation to measure the e-cloud must first be further developed, for instance e-cloud within dipole magnets cannot be accurately measured, although e-cloud is thought to accumulate at higher densities in magnetic field regions.

In order to properly measure beam halo, the dynamic range needs to increase to at least $10^6$ for 2D phase space measurements of hadron beams. The state-of-the-art is currently $10^5$ and has only been recently demonstrated. In addition, it has been shown that a 6D phase space distribution measurement is required to properly describe the full phase space structure of the beam. Such 6D measurements have been demonstrated but are not yet practical for operational use.

As new sources are developed that produce lower emittance beams, new high-resolution diagnostics are required in order to characterize and manipulate them.

### 4.3.4. Theme 4: Test stands to advance new technology and validate simulations.

As with simulations, test facilities for accelerator science and technology are a common thread required for advancing all four grand challenges. They are also important for student training.

**Research topics in Theme 4:**

Test stands for diagnostics and benchmarking
Test stands are being built and used to test novel mitigation methods for high intensity effects, and to develop advanced beam diagnostics and to benchmark simulation capabilities. These reside at both universities and laboratories. Test stands have the major advantage of unlimited beam time and freedom to make fast or high-risk configuration changes that would be limited in a production accelerator facility.

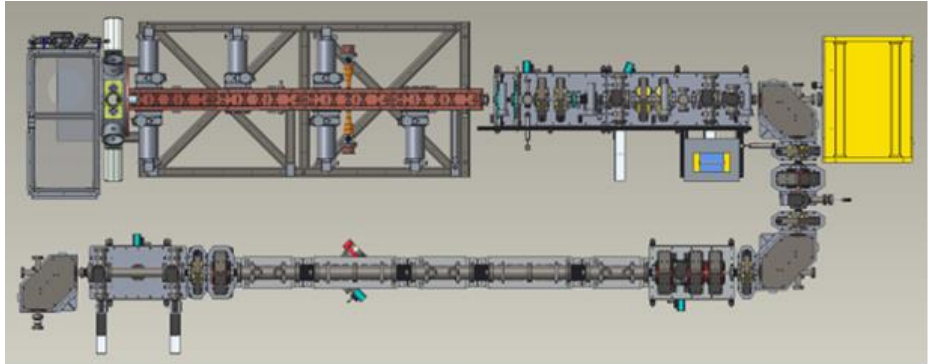
Beam Test Facility at SNS (image courtesy of ORNL).

Test facilities for the Intensity Frontier
Existing linac test stands for high intensity hadron beams include the FAST/IOTA facility at FNAL and the Beam Test Facility (BTF) at SNS. The IOTA ring test facility is focused on development of several high intensity concepts, including electron lens and integrable nonlinear optics. CBETA (Cornell) is equipped for studies of energy recovery and phenomena such as the beam-breakup instability, and optical stochastic cooling is under development at both IOTA (Fermilab) and CESR (Cornell). Research programs at these facilities are aimed at addressing limitations primarily at the Intensity Frontier.

Test facilities for extreme bunch compression and CSR mitigation
The FACET-II (SLAC), AWA (ANL) and ATF (BNL) test facilities are critical to development and validation of techniques for extreme beam compression and coherent synchrotron radiation mitigation for future colliders and accelerators.

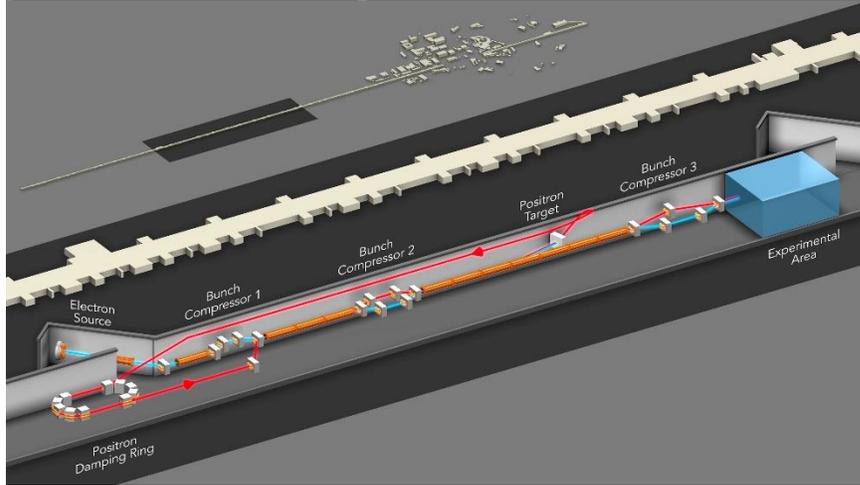
FACET-II facility schematic (image courtesy of SLAC).

## 4.4. WG4: Connections to other GARD roadmaps

WG4 focused on cross-cutting research topics that are related to existing GARD roadmaps and have an ABP component. The themes in this section are: (1) ABP connection to the Normal-Conducting Radio-Frequency (NCRF) roadmap [4]; (2) ABP connection to the Super-Conducting Radio-Frequency (SCRF) roadmap [4]; (3) ABP connection to the High-Field Magnets (HFM) roadmap [5]; and (4), ABP connection to the Advanced Accelerator Concepts (AAC) roadmap [6].

### 4.4.1. Theme 1: ABP connection to NCRF

**Research topics in Theme 1:**

Connection to roadmaps
The GARD RF Roadmap Strategy report was prepared in 2017 [4]. It defines decadal goals that deliver transformative impact for RF accelerator technology. The RF decadal integrated roadmap has two main components: advancement of RF structures and advancement of the RF power sources and auxiliary systems surrounding the structures. The overarching goal is to dramatically improve performance and cost by an order of magnitude or more. RF Accelerator R&D and Accelerator and Beam Physics R&D naturally connect on many fronts. One front is the exploration of wakefields and collective effects, e.g., heavily beam loaded multi-frequency structures, scaling of distributed-coupling accelerator structures to high energy, or experiments with small scale accelerators to enhance efficiency. Another front is the improvement of sources, e.g., electron sources (cryogenic rf guns, thermionic rf guns, THz guns), RF sources (optimizing efficiency and beam dynamics of advance concepts), or integrated modeling (merging beam dynamics and rf modeling for virtual prototyping).

Synergistic opportunities
Opportunities for synergies between NCRF and ABP involve topics directly related to NCRF concepts: wakefields and collective effects; RF gun and injector studies; field emission and multi-pactoring; RF system modeling. Other topics more related to NCRF applications include beam collimation and halos; space charge and CSR with short bunches; nonlinear beam optics for strong focusing systems; systematic tolerance and costing studies.

Simulations
There are, in particular, great opportunities for overlap on integrated modeling. Three examples of integration are: (a) integration of an RF field solver (ACE3P) with a beam dynamics code (IMPACT), enabling virtual machine analysis; (b) integration of ACE3P+IMPACT with a plasma code (Warp) for studying plasma processing used for enhancing cavity performance; (c) integration of ACE3P with a Particle-Matter Interaction Code (FLUKA), enabling Modeling of Radiation for Machine Protection. These integrations are enabled by the Consortium for Advanced Modeling of Particle Accelerators (CAMPA), an initiative funded by GARD (currently including LBNL, SLAC, FNAL and UCLA) and establishes a platform for cost effective application of accelerator codes.

### 4.4.2. Theme 2: ABP connection to SCRF

**Research topics in Theme 2:**

Synergistic opportunities
Superconducting radio frequency technology is a cornerstone technology for future high-energy particle accelerators. For high-intensity and high-brightness accelerators, the beam physics requirements strongly influence the design and parameters of the SRF acceleration cavities and cryo-modules, and vice versa. In particular, beam-cavity interactions involve cross-cutting activities between ABP and SRF:

- Beam instabilities and emittance dilution caused by beam-cavity interaction for high intensity beams and for beams with high phase space density (ABP Challenges 1 and 2).
- Careful analysis of these phenomena is to be an intrinsic part of the beam simulation tools for a "virtual particle accelerator" (ABP Challenge 4).
- Other effects related to the beam physics include dark currents caused by field emission (which may influence the accelerator operation), as well as Cooper pair disruption by EM fields of dense particle bunches.

Simulations and 'virtual particle accelerator'
One of the main cross-cutting tasks would include the development of a predictive "virtual particle accelerator" that includes the features of an SRF acceleration system. The virtual accelerator should allow the optimization of the design and operation regime, and enable determination of the technical and physical limitations. This entails the implementation of models for wakes in SRF acceleration systems (transition, transient, steady state, cavity components), high-order modes (HOMs) and dark currents.

### 4.4.3. Theme 3: ABP connection to HFM

**Research topics in Theme 3:**

Synergistic opportunities
Magnet technology is driving the cost and determining the reach of a future collider. Paradigm shifts that may influence or benefit from ABP-HFM interactions include: (a) significant reduction of the static multipole content constraint, (b) significant reduction of the magnetization constraints, and (c) large momentum acceptance optics. Modeling of magnet has advanced and accurately predicts 3D fields for "complex" magnet layouts. The field can now benefit from improved high-performance modeling tools to fully leverage the ABP-HFM linkages. For example, can magnet modeling be integrated into "virtual particle accelerators" for concurrent optimization of magnet designs with beam dynamics, RF, etc.?

### 4.4.4. Theme 4: ABP connection to AAC

**Research topics in Theme 4:**

Structure wakefield acceleration
There are strong connections between the ABP Grand Challenges and the AAC Roadmap [6]. Several Structure Wakefield Acceleration (SWFA) Roadmap items would benefit from ABP progress:

- Beam quality. Reduction of the transverse and longitudinal emittances of the main beam raise luminosity. The improvement of transverse and longitudinal emittances of the drive beam are key to propagation through the wakefield structures. In addition, the dimensions of structures scale with the frequency of the RF fields being generated, thus making the emittance requirements more demanding for higher frequency wakefields.
- Beam intensity. The drive bunch requirements of high charge and short bunch length generate very intense electron bunches
- Beam control. Wakefields excite both the short-range and long-range beam break-up (BBU) instability as SWFA operates in the strong wakefield regime.
- Beam control. The quest for high energy efficiency in the process of transferring energy from the drive bunches to the main bunches, makes it longitudinal shaping of both the drive and main bunch necessary. E.g. a reverse-triangle main bunch reduces beam-loading effects and double triangular longitudinal profile of the drive bunch allows for better utilization of the drive bunch energy in two-beam acceleration
- Beam control. In the collinear scheme of SWFA, the shaping of the drive bunches is also important, since it enables significantly higher transformer ratios.
- Beam quality. Robust beam optics is also a fundamental requirement in the propagation of drive bunches with fast decreasing energy, as they transfer their energy to the wakefields.
- Beam quality. Compact beam optics between stages of acceleration is also critical, since it directly affects the effective accelerating gradients that can be achieved.
- Beam Breakup (BBU) control is also a fundamental requirement for both drive and main beams.
- Beam prediction. High gradient and high energy efficiency wakefield acceleration lacks the ability start-to-end numerical simulations of long bunch trains composed of very short bunches. For example, a train of ~50 sub-picosecond bunches separated by ~1 nanosecond has physics operating on disparate scales, short-range single bunch effects (e.g. space charge, CSR, etc) combined with long-range wakefields.

Plasma Acceleration
Plasma accelerators offer the potential to deliver accelerating gradients of more than 10 GeV/m. In a plasma accelerator the power source is a highly energetic laser pulse or particle beam. The accelerating structure is created within the plasma, dynamically by the driver, on a pulse-by-pulse basis. Although there can be differences in behavior depending on the specific driver, much of the physics for the accelerated bunch is the same. In the case of beam-driven plasma-acceleration, most of the connections identified for the SWFA are relevant.

The 2016 Advanced Accelerator Development Strategy Report [6] outlines research roadmaps for plasma accelerators for the next decade. Strawman designs for high-energy colliders have been used to prioritize research programs that build on the current state of the art. Even at the conceptual level, it is clear that accessing the energy frontier will require plasma-based concepts to achieve high-gradient acceleration and doing physics at TeV-scale energies requires high beam quality and power efficiency to achieve sufficient luminosity. Thus, periodic conceptual integration, optimization, and maturity evaluation workshops are needed to maintain a strong connection between the AAC and ABP thrusts.

Synergistic opportunities
Along the way to these broader milestones, many accelerator and beam physics issues need to be pursued and understood. An incomplete list involves topics related to acceleration and emittance preservation. Issues related to achieving high-gradient, high-efficiency acceleration include: understanding the limits of beam loading that allows simultaneously high-efficiency, percent level energy spread and sub-micron emittance; longitudinal beam shaping to maximize the transformer ratio and minimize the number of stages; transverse shaping for quasi-linear regime acceleration and certain positron acceleration concepts; inter-stage optics designs that maximize the average gradient and concepts for high-gradient high-quality positron acceleration in plasmas. Issues related to preserving transverse beam quality (emittance) include: Coherent Synchrotron Radiation and inter-bunch correlation suppression; techniques for section-by-section alignment, correction and feedback; inter-stage focusing and dispersion control; applicability of plasma lenses; mitigation of multiple coulomb scattering, ion motion and mismatch; requirements on transverse and longitudinal drive/witness beam jitter. Because the plasma response is not symmetric for electrons and positrons, new concepts need to be developed for tailoring some combination of the drivers, plasma source (density profile) and the accelerated positron beam to achieve the requisite high-gradients and beam quality.

Many of the items listed above address elements of the ABP Grand Challenges but are not yet pushing to the extremes mentioned.

Grand Challenge #2: Beam quality – plasma accelerators are pushing on brightness and quality preservation during acceleration and hope to demonstrate orders of magnitude improvement in the next few years compared to state of the art parameters.

Grand Challenge #3: Beam control – longitudinal beam shaping for high-transformer ratios and better beam loading; transverse shaping for accelerated beams in LWFA collider concepts in quasi-linear regime and certain positron acceleration concepts.

Grand Challenge #4: Beam prediction – there is a need for better code integration between beams and plasma PIC codes; beam codes need validation of CSR models (hosing seed) and effects at low emittance and high-peak currents; reduced models work well when appropriate (e.g. quasi-static in QuickPIC, HighPACE) but ultimately codes that scale to Exascale for modeling multiple stages and parameter scans (Warp-X Exascale development) will be required.

## 5. SUMMARY from WORKSHOP #2

**HEP GARD Accelerator and Beam Physics: Community-driven Strategic Roadmap Workshop #2 (FNAL, April 15-16, April 22, April 30, May 9, May 21, 2020)**

The goals of the workshop [3] were to collect community input in identifying the key Accelerator and Beam Physics (ABP) needs to fulfill OHEP's GARD mission and to develop a decadal roadmap for thrust activities that OHEP could support.

Workshop #2 focused on research areas for each of four working groups:

**WG5:** Advanced accelerator instrumentation and controls.
**WG6:** Modeling and simulation tools; fundamental theory and applied math.
**WG7:** Early conceptual integration and optimization, maturity evaluation.
**WG8:** Connections and synergies with non-HEP accelerator science activities.

### 5.1. WG5: Advanced accelerator instrumentation and controls

WG5 explored potential roadmaps for instrumentation and controls beyond today's state-of-the-art in support of the ABP Grand Challenges (GC). Instrumentation and controls enter most of the GCs and are needed to support the generation and characterization of beams over a wide range of parameters (from quantum-degenerate electron beams to high-intensity hadron beams). The themes in this section are: (1) Control; (2) Measurement; and (3) Prediction.

#### 5.1.1. Theme 1: Control

The Control theme concerns the development of output instrumentation (e.g. power supply, magnet) and algorithms (e.g. singular value decomposition, ML)) used to set a parameter (e.g. beam position) to a desired set-point. Instruments are an integral part of the feedback system and are desired to have wide dynamic range and low latency coupled with high precision and accuracy. Given the complexity of large-scale accelerators, with millions of process variables, the set-points of the parameters via machine-learning (ML) has recently become an active area of research. Our roadmap will identify the beam or machine parameters for which we seek improved control in the upcoming decade.

Research topics associated with the control theme:

- RF phase/amplitude stability
- RF-laser synchronization at the sub-fs level.
- Feedback to mitigate collective effects (space charge, multi-bunch interaction,…)
- ML/AI-based control of accelerator parameters

### 5.1.2. Theme 2: Measurement

The Measurement theme deals with developing diagnostics and techniques to characterize the beam, ranging from its scalar macroscopic properties, phase-space distribution (either by direct observation or indirectly) and, ultimately, the location of each particle within the bunch. This effort supports the grand challenge of tailoring a bunch at the single-particle-level.

Research topics and milestones associated with the measurement theme:

- Measure betatron functions of hadron beam with sub-% accuracy;
- Measure mega-Ampere peak-currents;
- Measure ultra-low emittance at the Heisenberg-limit from source or after cooling;
- Measure temporal beam shapes with precision beyond the femtosecond regime;
- Measure beam halos with a large dynamical range in 6D: Halo can be measured with a high dynamic range in 1D but the dynamical range decreases with the dimensionality of the phase space [Note: data analysis for high volume, high dimensional data (e.g. 6D) is difficult; this should be coordinated with WG2.]
- Measure beam spin-polarization with high accuracy and speed;
- Measure 6D (or projected) phase-space-distribution diagnostics capable of measuring complex correlations within the bunch;
- Improve diagnostics usability: single-shot, non-destructive, wide dynamic range, etc;
- Develop simultaneous measurement of multiple beams (multiple species as in electron/hadron cooling section, or multiple same-species beams at different energies e.g. recirculating or energy-recovery linacs).

### 5.1.3. Theme 3: Prediction

The Prediction theme entails the development of methods that can use observable signatures to predict the state of the accelerator or beam and possibly use this information to control the accelerator (feedback). This theme relies on the development of ML-based virtual diagnostics that can be used to infer non-directly observable quantities.

Research topics associated with the prediction theme.

- Automatic tuning;
- Online modeling using simulations augmented with measured data;
- Anomaly detection, using machine learning to predict possible failures (e.g. resonator trip) and guide preventive action;
- Development of advanced (virtual or improved) beam diagnostics;
- Development of advanced ML-based feedback systems;
- Develop better physics-informed models (e.g. proper modeling of fringe fields, etc...);
- Will ML demand higher precision of accelerator models, input measurements and output controllers or will ML relax requirements of all three?
- Explore whether ML can be productively applied to optimization of beam polarization, beam halo, etc.

## 5.2. WG6: Modeling and simulations tools: fundamental theory and applied math

WG6 focused on the current state-of-the-art, needs, visions and possible roadmaps in theory and simulation tools. The themes in this section are: (1) Analytical methods; (2) Modeling of intense beams; and (3), Numerical tools, collaborative frameworks, and virtual accelerators.

### 5.2.1. Theme 1: Analytical methods.
Analytical properties of nonlinear dynamics in accelerators are far from completely understood. The community must develop more advanced tools in order to meet future requirements on beam brightness and quality. Two major directions are the use of analytical methods to develop conceptually new machines and tools to optimize or analyze existing accelerators.

**Research topics in Theme 1:**

Search of new integrable systems

There is no general method of finding integrability conditions for dynamical systems with many degrees of freedom. Usually, we discover one by luck or by introducing sufficient symmetries. We should continue to search for these and possibly implement them in accelerator systems. Examples include round colliding beams, IOTA, use of electron lenses. New directions should be explored as well. We can move from integrable to near integrable, perhaps chaotic but bounded (for example systems with attractors), or even structurally stable analytical models.

General theory

- Development of perturbation techniques for quick estimates of dynamical aperture and dynamical variables such as fundamental frequencies is a direction in which we must proceed.
- Develop a new modern language and methods for the description of nonlinear optical functions in a machine.
- Develop new stochastic and PDE approaches for Spin-Orbit dynamics in addition to the development of simulation tools. Possible applications for EIC, FCC-ee, CEPC and etc. have to be considered.
- Open theoretical questions have to be addressed; in particular better understanding of correction terms to Derbenev-Kondratenko formulas and a deeper investigation of Markov approximation of single-particle orbital motion due to synchrotron radiation.

Incorporation of AI (including Machine Learning based on adjoint methods or other)

- Using AI to search for integrable systems could proceed in two possible directions: (1) recognition of integrable or near-integrable systems, and (2) development of algorithms that directly search for one.
- Use of AI for tuning or optimizing an existing lattice (or to help in the design of future machines) is different from using AI for operational optimization. In this case, analytical models should be used as a system to optimize.

### 5.2.2. Theme 2: Modeling of intense beams

Despite recent progress in understanding halo formation and mitigation in intense beams, there remain many unaddressed fundamental questions, such as whether halo-free beam equilibria exist, if it is possible to inject beam without creating halos, and if it is possible to generate beams that can be injected without halos. Substantial development is needed in the modeling of intense beams in order to answer these questions.

The exploitation of nonlinear elements for Landau damping (e.g. HL-LHC, IOTA, PIP-III) calls for more accurate and efficient tracking algorithms in highly nonlinear or poorly characterized fields.

Numeric noise in space charge simulations could drive unphysical diffusion, which leads to incorrect prediction of beam behavior in lattices based on integrable optics. Understanding and controlling such effects is an important goal.

Coherent synchrotron radiation (CSR) in intense, short electron beams can severely limit the beam quality. While there have been a number of codes that implement 1D, 2D, or 2.5D models, there are currently no codes that can accurately and reliably model 3D CSR effects.

**Research topics in Theme 2:**

Modeling of Intense Beams

- Characterization of nonlinear applied fields using realistic Maxwellian models and using them for symplectic tracking. Development of algorithms to extract a concise, smooth representation of the vector potential from boundary data for highly nonlinear elements (without resorting to a Taylor expansion) to use in symplectic tracking for a realistic ring.
- Self-consistent matching of intense beams in a nonlinear lattice and studying the sensitivity to the initial phase space distribution is a continuing goal. Recent work on beam equilibria in nonlinear lattices could be extended to the case of intense beams in a general nonlinear, s-periodic lattice to produce a numerical tool for the generation of matched beams. Other methods of generating integrable lattices, along with studies of their robustness, could be explored.
- Understanding numerical particle noise in intense beams and its role in driving diffusion in nonlinear lattices and mitigating such effects is of ongoing interest. It may be profitable to apply techniques from statistical mechanics and kinetic theory to study properties of a near-integrable, nonlinear N-body Hamiltonian system of simulation particles. Improved smoothing and symplectic integration could reduce noise.
- Design the transport for the injection into a ring (e.g. IOTA) that results in matched beams (including the effect of a kicker) or the generation of a matchable beam. It is desirable to design the transport to match a given beam, and if that is not possible, perhaps the beam could be designed to match the lattice. Adjoint methods and machine learning could be used.
- Develop 3D full PIC simulation for intense beams. Improved computational power has made this feasible. Scale disparities and boundaries must be dealt with effectively. Also of interest are symplectic, high-order PICs, computational algorithms on GPUs, and high-accuracy solvers.

3D CSR Theory and Simulation Code

- Develop a 3D CSR theory that describes the full physics involved in the phenomenon. Explore both the time domain and the frequency domain. Progress in this area could help determine strategic placement of shielding.

- Developing a 3D simulation code that can accurately and efficiently model the physical process, including transient effects, for extremely bright electron beams in accelerators. Such a simulation should accommodate a quasi-realistic vacuum chamber, arbitrary phase space distributions, be self-consistent, have the ability to handle ultra-short (sub-micron) bunches, and have a sufficiently fast turn-around.

### 5.2.3. Theme 3: Numerical tools, collaborative frameworks, and virtual accelerators

No major accelerator project can proceed without being thoroughly modeled by comprehensive, detailed simulations. There is a need (in addition to fast, reduced models) for full physics 6D computer simulations (often based on the Particle-In-Cell method). Although there are many beam and accelerator codes in existence (developed independently, with licenses ranging from fully open to fully closed source) prediction of halo or emittance growth to even within a factor of two is challenging. Furthermore, simulations often only partially address the accelerator and associated beam physics, sometimes missing important couplings of effects. Without full system modeling, opportunities are missed for system optimization and for achieving the maximum benefit from machine learning.

There is thus the need for:
- Further development of algorithmic and High-Performance Computing.
- Collaborative frameworks that run on desktop computers, supercomputers, in the cloud.
- Integrated, multi-physics, self-consistent, start-to-end simulations that can predict the behavior of beams in accelerators "as designed/built".
- For all these tasks, the use of machine learning should be explored.

**Research topics in Theme 3:**

Numerical tools

- High performance parallel computing is essential to predict halo and emittance growth accurately, which is beyond the current state-of-the-art. There is thus a need to push for faster, more accurate algorithms and codes.

- Computing hardware and technology are undergoing a generational change (e.g., CPU to GPU), and simulation codes must be upgraded to efficiently execute on new platforms. The question arises whether the Fortran legacy is an asset, a drawback, or something neutral.
- Operating systems, compilers, computing hardware, programming languages and the availability of people fluent in them changes over time. In spite of this, software must be maintained so as to remain viable in the long-term. Codes have been lost because they became unmaintainable. Their capabilities have had to be re-implemented.
- Explore how machine learning can be used to achieve better algorithms, more accurate and faster codes.

Collaborative frameworks

- Develop collaborative framework(s) to organize code development efforts toward integrated, multi-physics, start-to-end modeling. To this end, standardization of output data and input scripts should be developed, and their adoption encouraged. It is essential to enable the coupling of different codes and to have standards for uniform data. In this way machine learning capabilities could be leveraged. The openPMD standard is already widely used for particle and field data from simulations and experiments.
- Build on efforts to develop online framework(s) in the Cloud (e.g., Sirepo) to enable integrated, collaborative research (with computational reproducibility) and education.
- Note that a mix of licensing is possible, but degree and effectiveness of collaborations increase with the level of openness of the various components. Although open source codes are not required, they are favored.

End-to-end Virtual Accelerator (EVA) – GC#4

- Develop End-to-end Virtual Accelerators (EVA) that can predict the behavior of beams in accelerators "as designed/built". Like a flight simulator, an EVA should allow modeling operation of the accelerator in "real time". Machine learning could be used to develop fast virtual models.
- Build on the framework task to connect codes and frameworks for integrated, multi-physics, self-consistent, start-to-end simulations that include "everything". Examples of some things important to include are accurate models of accelerator components, collective effects, dynamic machine configurations, and energy deposition.

- specialized tools are needed to model structured plasmas for next generation accelerators (MHD or Vlasov-Fokker-Planck). Modeling needs are expected for plasma channels for acceleration, plasma lenses for focusing, and plasma afterglow diagnostics.

## 5.3. WG7: Early conceptual integration and optimization, maturity evaluation

WG7 focused on a review of research needs and opportunities in GARD that are oriented to improving existing complex accelerators, developing new concepts for future accelerator facilities, and possible significant upgrades of existing machines. The emphasis was on the conceptual integration and optimization of accelerator physics constraints and engineering technical challenges needed to arrive at a desired overall conceptual design.

The themes in this section are: (1) Accelerator physics topics for the near term (<10 years), and (2) Accelerator physics problems for long term accelerator facility plans (>10 years).

### 5.3.1. Theme 1: Accelerator physics topics for the near term (<10 years)

Some accelerator topics are relevant for the near term (<10 years), for example, those related to well established facility projects with CDRs/TDRs. They do not strongly rely on GARD for the present design choices or performance projections but could benefit from future GARD ABP R&D that may result in either performance enhancements, cost risk mitigation, or shorter commissioning period.

**Research Topics in Theme 1:**

High Power Proton Sources (1 MW – multi-MW)

- Beam physics issues related to beam loss control (space-charge, instabilities, collimation, e-lens compensation, integrable optics, etc) will benefit from innovative approaches, theoretical and experimental studies (at e.g. IOTA, and operational accelerators in the US and abroad) and validated computer models/codes. A key challenge would be to reduce particle losses (dN/N) at a faster rate than increases in achieved beam intensity (power) (N).
- Expanded small topical national and international collaborations could prove quite successful and useful, as well as collaborative work synergistic with the goals of EIC, MC, NuSTORM and ADS.

Circular e+e- colliders (FCCee, CepC, gamma-gamma Higgs factory)

- Several new developments call for expansion of general studies of optimized beam and beam-beam parameters for circular Z-W-Higgs-Top factories including 3D beam size flip-flop from the beam-beam effect, polarization and IR collision optimization in a collider.
- An Interaction Region (IR) design with gamma-gamma laser-beam conversion should be performed, in parallel with possible design considerations of the corresponding high-power laser system.
- Pico-meter vertical emittance preservation techniques in high-charge circular colliders with strong focussing IR, detector solenoids, and beam-beam effects (in synergy with SuperKEKB).

Linear e+e- Colliders (ILC, CLIC, LowT-NCLC)

- To reduce the expected commissioning time of linear colliders, end-to-end emittance preservation simulations (including parallel processing) as well as tuning tools (including Machine Learning and Artificial Intelligence) for linear colliders should be developed. Experimental tests of the beam-based alignment techniques in the presence of realistic external noise sources are needed and possible at high energy linac-based facilities such as XFEL, LCLS-II, and FACET-II.
- Novel new techniques for linear colliders, such as a plasma-based final focus or a cryogenic normal conducting RF linac design, need to be evaluated and advanced through comprehensive beam physics studies performed in tandem with facility design and cost analysis.

Hadron Colliders (FCChh, SppC, HE-LHC)

- Accelerator physics issues for vacuum system designs with electron cloud interactions in TeV hadron colliders with bunch spacing less than 25 nsec.
- Over the next decade, many valuable accelerator physics explorations can be done at CERN, RHIC, IOTA, and other accelerators on topics of importance ranging from more efficient collimation techniques, to electron lenses, to dynamic aperture optimization methods.
- Magnet design studies aimed at higher fields, cost reduction, and better field quality, especially for lower injection energy or with possible new integrable optics solutions.
- Studies to obtain lower emittances from new particle sources for injecting beams in high-bunch-charge colliders.
- Exploration of lower cost hadron main colliding rings by using top-up injection.

### 5.3.2. Theme 2: Accelerator physics problems for long term accelerator facility plans (>10 years)

Some accelerator physics topics are relevant for long term accelerator facility plans (>10 years), such as those with intermediate readiness and others close to "strawman" machine designs. These include or may include advanced accelerator concepts, ERL-based designs, or innovations resulting in lower wall plug power; these advancements are crucial to making those accelerators scientifically, technically, and fiscally feasible.

**Research Topics in Theme 2:**

Superbeams 3-10 MW (PIP-III) and Neutrino Factories

- Beam physics and design optimization studies towards conceptual design of 3-10 MW superbeams facility design (focusing on power efficiency and cost per physics result outcome).
- Optics/Dynamic Aperture methods needed (integral nonlinear optics, Vertical-FFAG, etc) to increase the beam lifetime in racetracks of NuFact.
- Very-fast-ramping and high-field radiation-hard magnets (expanding on the US MDP); very high-power tunable RF (expanding on the GARD RF roadmap), laser stripping injection schemes.

Muon Collider and Neutrino Sources (Higgs-3-14 TeV MC)

- Design optimization studies toward a scientifically, technically, and fiscally possible muon collider – ideally, via joining the world muon effort, aimed at the CDR in 5-7 years (and TDR in 10-15 years).
- Studies of new and improved muon emittance cooling mechanisms – from 6D cooling to positron ring-based muon sources. Final stage muon cooling studies are needed.
- Explore challenges and opportunities of orders of magnitude higher muon production rates.
- Accelerator protection from decaying muons and neutrino radiation hazard mitigation.
- Very-fast-ramping and high-field radiation-hard magnets (expanding on the US MDP).

Advanced Concept Colliders (Beam-, Laser- Plasma, DWA, microstructures)

- New collider concepts with overall comprehensive design optimization and systematic accounting of all beam physics and technology related issues. For example, these are needed to progress the AAC collider optimization beyond current "strawman design" status. These studies should be coordinated with concurrent conceptual development of detectors.

- Optimized AAC electron acceleration technology for a collider; optimized positron acceleration; plasma multi-cell layout optimization, and the physics of drive beam instabilities and optimization.
- Optimized beam power to wall-plug power efficiency.
- Overall cost reduction, lifetime studies, and reduction of component damage are needed for the AAC colliders.

### 5.4. WG8: Connections and synergies with non-HEP accelerator science activities

WG8 focused on exploring the synergies of GARD accelerator and beam physics with interests from other offices of science [7]. After presentations that covered topics in nuclear physics, basic energy sciences, national security, and the advanced scientific computing research, the conclusion was that ABP synergies are broad, and essentially everything is cross-cutting. All four ABP Grand Challenges apply to all categories of research machines. The themes in this section are: (1) Some ABP resources require universal support from the accelerator science and technology community; (2) Methods to meet some ABP challenges are relevant to all classes of research accelerators; (3) Methods to meet some ABP challenges are relevant to more than one class of research accelerator; and (4), Some research of importance to only one class of accelerator may later become relevant for another class of accelerator.

#### 5.4.1. Theme 1: Some ABP resources require support from the US accelerator science and technology community, for example in

- Education of next generation scientists and engineers
- Facilities dedicated to testing accelerator concepts
- Development of theoretical and computational/machine learning tools

**Topics in Theme 1:**

Education of the next generation

At present education of the next generation is formally supported by the USPAS, the DOE Traineeship program, university-based research including the NSF-supported Center for Bright Beams, , some support from national labs (for example, the graduate program in accelerator physics at FNAL), undergraduate research experience such as the Lee Teng Internship, and some support from small SBIR-based companies.

Going forward, it is critical to maintain support for the existing educational mechanisms. Enhancement of existing mechanisms is an important objective. Giving awards to proposals that are most likely to strengthen these mechanisms with either new innovations or new cross-connections would be one way to do this. Strengthening support of student-oriented services at conferences would be another way. Other measures could be considered, such as supporting a strong central database with statistics on the number of graduates into the field each year, and information on the path that was taken to get there. A stronger centralized information system regarding job opportunities, internships, and educational opportunities, across industry, academia and the national labs would also strengthen the pipeline of new professionals.

Facilities dedicated to testing accelerator concepts

Currently there are several dedicated accelerator research facilities, such as FACET (SLAC), AWA (ANL), ATF (Brookhaven), and IOTA/FAST (FNAL). Unique capabilities are also available at CBETA at Cornell, and at photoinjectors at UCLA, Cornell and the national labs. Such facilities are invaluable for pushing new accelerator concepts and technology.

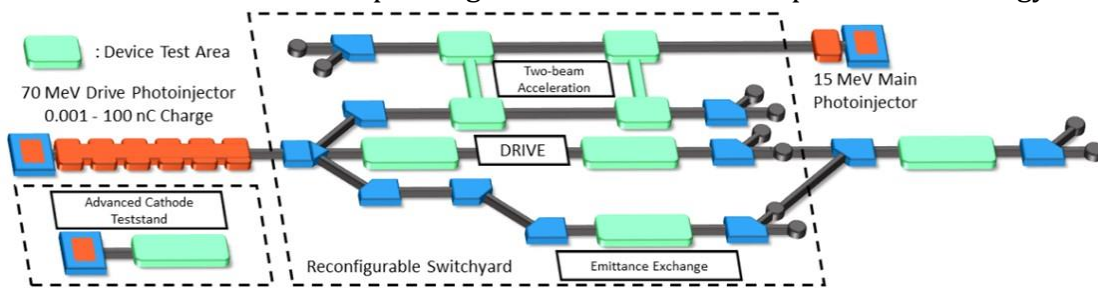

Argonne Wakefield Accelerator (AWA) schematic (image courtesy of ANL)

Going forward, it is critical to maintain and enhance support for accelerator test facilities, as long as they remain productive research facilities. Enhancement could be encouraged with awards to proposals that are most likely to strengthen these facilities with new innovations or cross-connections, especially as they show promise for making progress on the grand challenges.

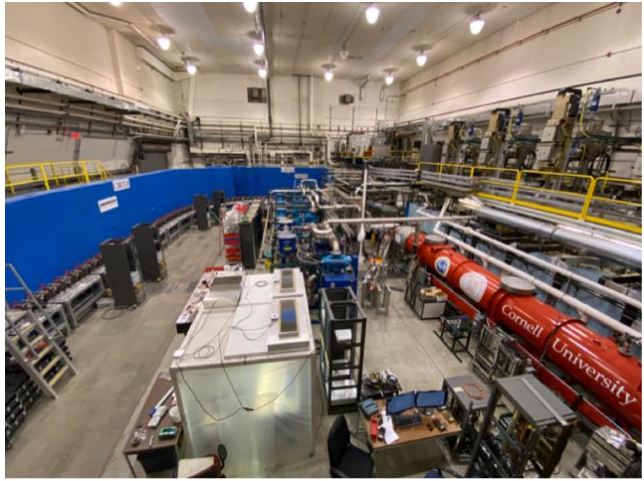

The Cornell-BNL ERL Test Accelerator (CBETA) on Cornell's campus (photo: Cornell).

Development of theoretical and computational/machine learning tools

There are multiple efforts in the community to make progress in these areas. It is important to integrate these efforts to the extent possible, see 'Collaborative frameworks' under WG2 Theme 3. Support for effective collaborative frameworks should be enhanced in order to improve standardization and organize code development efforts. The development of online frameworks for collaborative research and education should be supported. Support should be especially encouraged for efforts that are most likely to strengthen collaborative frameworks, generate new innovations, or establish new cross-connections, especially as these show promise for making progress on the grand challenges.

WG4 discussions regarding accelerator modeling, measurement, machine learning (ML) and controls, had some of the following main takeaways: (a) ML can benefit from beam physics and more cross talk is needed; (b) measure first and then control, measurement is easier and can be put to novel use in the case that control is not possible; (c) use of ML needs better diagnostics with improved accuracy and resolution, and ML can also be used to develop virtual diagnostics; (d) parallel multi-objective optimizations, accurate linear and nonlinear modeling are needed; and (e) it is important to achieve strict correspondence between what is designed and what is built.

**5.4.2. Theme 2: Methods to meet some ABP challenges are relevant to all classes of research accelerators**

Some items required for meeting the challenges of ABP are important for all classes of research machines, and advances in these areas would likely be productively shared for modification and adaptation.

**Research Topic in Theme 2:**

All (or most) accelerators share certain cross-cutting issues. Building energy efficient machines is an important issue and is key to future development. More attention must be paid to this going forward from the single components level up to overall facilities. There are many important beam dynamics considerations that span all classes of machine such as beam stability, phase space manipulation, injection/extraction design, gas scattering, and so on. Beam compression dynamics are not fully understood, and progress on this would benefit multiple classes of machine.

### 5.4.3. Theme 3: Methods to meet some ABP challenges are relevant to more than one class of research accelerator

Some critical research, while not of universal concern, has overlap between certain classes of machines. Research could be productively shared between different groups of stakeholders (for example HEP machines and light sources).

**Research Topics in Theme 3:**

<u>Electron-Ion Colliders</u>

Electron-ion colliders (EIC), have cross-cutting accelerator physics conceptual similarities and challenges with HEP colliders. Collision regions have similar issues such as beam-beam tune shift, crossing angle and crab crossing issues, and interaction region design. The storage rings have similar issues such as beam lifetime, intrabeam scattering, and the potential need for beam cooling.

<u>Light Sources</u>

Light sources also have cross-cutting accelerator physics conceptual similarities and challenges with HEP machines. The challenge of small alignment tolerances needed for fourth-generation storage rings has similarity to some alignment challenges faced by HEP colliders. Both types of machine have radiation damage concerns, require low emittance, benefit from ultrafast diagnostics and precision timing, require advances to achieve better vacuum, and benefit from high-gradient acceleration.

<u>NNSA and Security</u>

Many accelerator projects exist in the NNSA domain. Better information exchange and collaboration would be beneficial for the GARD ABP program.

### 5.4.4. Theme 4: Some research important to only one class of accelerator may later become relevant for another class of accelerator

Some research, while not of immediate concern for a given stakeholder (HEP), may become critically relevant in the future. This may include advancements for another type of research machine, or new ideas that have not yet matured. It is important to have a mechanism for supporting new, perhaps unforeseen thrusts.

**Research Topics in Theme 4:**

ERLs are not currently used for HEP machines but are relevant to some schemes of future colliders.

Preservation of spin dynamics is now important primarily for EIC machines but may become increasingly important in future HEP machines, for example at the ILC.

## 6. CONCLUSION

An effort to gather community input to prepare for the development of the Accelerator and Beam Physics (ABP) Roadmap by the HEP General Accelerator R&D (GARD) program has been completed. Two preparatory workshops were held, an in-person workshop at LBNL in December of 2019 and a virtual workshop hosted by FNAL and ANL in April-May of 2020. These workshops convened university and laboratory scientists active in ABP research to develop an R&D roadmap for the long-term vision for enabling future DOE HEP capabilities. The workshops were organized around four grand challenges (GCs): *Intensity*, *Beam Quality*, *Beam Control* and *Beam Prediction*. Research topics with milestones were established for each of the GCs. This report documents the outcome of these workshops and is made available to the Office of HEP in anticipation of the HEP workshop to establish their ABP Roadmap for the GARD program.

## 7. ACKNOWLEDGMENTS

This work has been supported and/or co-authored by
- Fermi Research Alliance, LLC under Contract No. DE-AC02-07CH11359 with the U.S. Department of Energy, Office of Science, Office of High Energy Physics;
- UT-Battelle, LLC, under contract DE-AC05-00OR22725 with the U.S. Department of Energy;
- LBNL under Contract No. DEAC02-05CH11231 with the U.S. Department of Energy, Office of Science, Office of High Energy Physics;

[7] Basic Research Needs Workshop on Compact Accelerators for Security and Medicine: Tools for the 21st Century, May 6-8, 2019, https://doi.org/10.2172/163112138